\documentclass[twoside]{article}

\usepackage{lipsum} 

\usepackage[sc]{mathpazo} 
\usepackage[T1]{fontenc} 
\usepackage{microtype} 

\usepackage[hmarginratio=1:1,top=32mm,columnsep=20pt]{geometry} 
\usepackage{multicol} 
\usepackage[hang, small,labelfont=bf,up,textfont=it,up]{caption} 
\usepackage{booktabs} 
\usepackage{float} 
\usepackage{hyperref} 
\usepackage{natbib} 
\usepackage{graphicx}
\usepackage{caption} 
\usepackage{subcaption} 
\usepackage{lettrine} 
\usepackage{paralist} 

\usepackage{abstract} 

\usepackage{titlesec} 
\renewcommand\thesection{\Roman{section}} 
\renewcommand\thesubsection{\Roman{subsection}} 
\titleformat{\section}[block]{\large\scshape\centering}{\thesection.}{1em}{} 
\titleformat{\subsection}[block]{\large}{\thesubsection.}{1em}{} 

\usepackage{fancyhdr} 
\title{\vspace{-15mm}\fontsize{24pt}{10pt}\selectfont\textbf{Detecting the Ultimate Power in the Universe with LSST}} 

\author{
\large
\textsc{Michael B. Lund}\\
\normalsize Vanderbilt University \\ 
\normalsize \href{mailto:michael.b.lund@vanderbilt.edu}{michael.b.lund@vanderbilt.edu} 
\vspace{-5mm}
}
\date{}

\begin{document}

\maketitle 

\thispagestyle{fancy} 


\begin{abstract}

\noindent Large time-domain surveys, when of sufficient scale, provide a greatly increased probability of detecting rare and, in many cases, unexpected events. Indeed, it is these unpredicted and previously unobserved objects that can lead to some of the greatest leaps in our understanding of the cosmos. The events that may be monitored include not only those that help contribute to our understanding of sources astrophysical variability, but may also extend to the discovery and characterization of civilizations comprised of other sentient lifeforms in the universe. In this paper we examine if the Large Synoptic Survey Telescope (LSST) will have the ability to detect the immediate and short-term effects of a concave dish composite beam superlaser being fired at an Earth analog from an alien megastructure.

\end{abstract}


\begin{multicols}{2} 

\section{Introduction}
\lettrine[nindent=0em,lines=3]{S} ome of the most significant discoveries in astronomy have been those that have occurred incidental to observations that had other objectives. Quasars were initially discovered as radio sources without optical components during all-sky radio surveys \citep{Schmidt1963}. During tests of a radio antenna, unexpected static was discovered that would later be explained as the leftover signal from the Big Bang \citep{Penzias1965}. Unusual radio pulses on the order of one per second were tongue-in-cheek coined 'LGM1', referring to 'Little Green Men', before being better explained as pulsars \citep{Hewish1968}. Detectors that were intended to monitor nuclear launches from Earth discovered other signals that came from sources outside the Earth, now known as gamma ray bursts \citep{Klebesadel1973}.

Other observations continue to hold on to an aspect of mystery. The famed 'WOW!' signal observed by SETI in 1977 possessed many of the characteristics that had been expected of an intentional radio message. Forty years after it was initially received, and despite further research, the true nature of this signal is still unknown \citep{Gray1994}. It still remains a candidate for extraterrestrial communication or some astrophysical event that we are not yet aware of. Much more recently, the observations of KIC 8462852 by \emph{Kepler} have provided an extremely unusual variable signal \citep{Boyajian2016}. Some of the suggested interpretations of this phenomenon have been astrophysical in nature, however there has also been speculation and follow-up observations that have framed this in the context of alien civilizations \citep{Harp2015, Wright2015a}.

In this paper, we look directly at the ability of telescopes to observe a photometric event that would be indicative of potential extraterrestrial activity. In doing this, we focus on a very specific test case, a concave dish composite beam superlaser as part of an alien megastructure being used to destroy an Earth-like planet. We further limit the scope of our consideration here by primarily addressing this question in the context of what will be possible with the Large Synoptic Survey Telescope (LSST), but this could easily be broadened to include additional ground- and space-based facilities.

\section{Large Synoptic Survey Telescope (LSST)}

\begin{figure*}[!htb]
  \begin{center}
   \includegraphics[width=\textwidth]{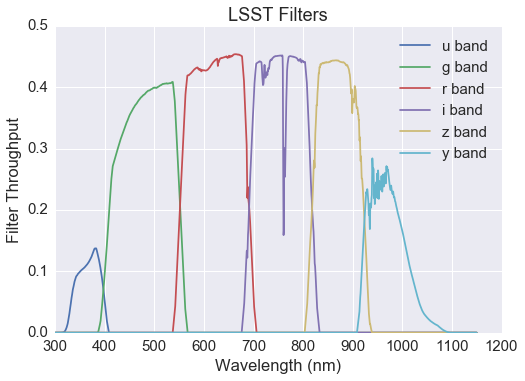}
  \end{center}
  \caption{The throughputs (including atmospheric effects) for the six bands that LSST will observe in.}
  \label{fig:filters}
\end{figure*}

The Large Scale Synoptic Survey Telescope (LSST) is an 8.4-meter telescope currently being constructed in Chile \citep{Ivezic2008}. With first light anticipated in 2020, LSST will spend ten years observing the entire southern sky in six photometric bands \emph{ugrizy}, with sensitivity from 16th to 24th magnitude. While a large number of potential science results have been examined previously, the matter we discuss in this paper seems to have not received any consideration as of yet \citep{LSSTScience2009}. Many of the observational parameters for LSST have not been finalized, but for purposes of this paper we use the LSST filter throughputs that have been defined thus far\footnote{https://github.com/lsst/throughputs/tree/master/baseline}. The throughputs of all six bands are shown in Figure~\ref{fig:filters}.

\section{Blast Modeling}
We investigate the activity and immediate aftermath of a planet-destroying laser blast with a series of approximations. We consider our target to be an Earth-analog, and so we use the properties in Table~\ref{table:earth} for this planet, with values from \citet{Kite2009}. We additionally use approximations of the average temperature of the core and mantle as 6000K and 1270K, respectively. Previous work has already examined the question of the energy needed to destroy a planet in this way, and we use their value of $2*10^{32}$ J in order to destroy an Earth-like planet \citep{Boulderstone2011}. However, it would not be realistic to treat the super-weapon as fully efficient, and so we use values based off of nuclear explosions, where 50\% of the energy goes into the kinetic energy of the planet, 35\% into thermal radiation that raises the temperature of the planetary material, and 15\% into an immediate, short-duration flash of electromagnetic radiation\footnote{https://www.remm.nlm.gov/nuclearexplosion.htm}. The observable energy from the explosion then comes from two components, the immediate release of energy during the explosion (what we refer to as the 'flash') and the long-term thermal radiation from the debris of the planet (what we refer to as the 'remnant'). For the flash, we treat this as a blackbody with a surface of the Earth that will release all of the energy of this component in 2 seconds, or the equivalent of blackbody radiation for a surface at $10^{6}$ K. We consider the debris of the planet to be well-mixed and be of a single temperature, and when this is calculated for the total energy, we find it to be a blackbody with a temperature of 29,000K. As this is occurring while the planet is being destroyed, the radius will be increasing, however as the escape velocity is 11 km/s we treat this object as consistent with earth-sized for the immediate aftermath. A more time-dependent examination would require accounting for the debris cloud growing in size, as well as the cooling of the debris (a time scale on order of 100 days if approximated as linear cooling) and changes to the optical depth of the debris cloud.

\begin{table}[H]
\caption{Earth Properties}
\label{table:earth}
\centering
\begin{tabular}{lll}
\toprule
Parameter & Value & Units \\
\midrule
Earth mass & $5.97*10^{24}$ & kg \\
Earth radius & $6.37*10^{6}$ & m \\
Core mass fraction & 0.325 & \\
\shortstack[l]{Specific heat\\ capacity, mantle} & 914 & J K$^{-1}$ kg$^{-1}$ \\
\shortstack[l]{Specific heat\\ capacity, core} & 800 & J K$^{-1}$ kg$^{-1}$ \\
\bottomrule
\end{tabular}
\end{table}

\begin{figure*}[!htb]
  \begin{center}
   \includegraphics[width=\textwidth]{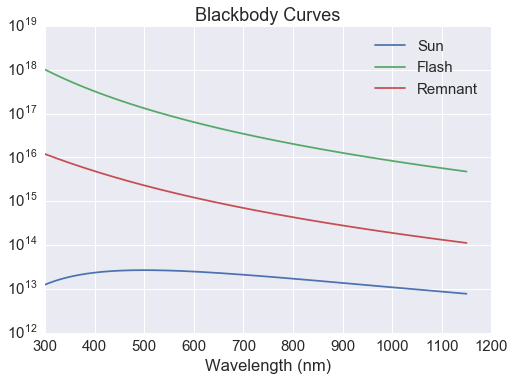}
  \end{center}
  \caption{Blackbody curves for the initial flash of the explosion as well as the debris remaining afterward. The blackbody curve for a solar mass star is included for comparison.}
  \label{fig:blackbodies}
\end{figure*}

We show the blackbody curves for the flash and the remnant in Figure~\ref{fig:blackbodies}. We also include a blackbody curve for a Sun-like star at 5800K for comparison. We then convolve each of these blackbody curves with the filter throughputs for LSST. Unsurprisingly considering the high temperatures involved, we see that the most significant contributions from both the flash and the remnant will occur in the bluer bands.

\begin{figure*}[!htb]
  \begin{center}
   \includegraphics[width=0.45\textwidth]{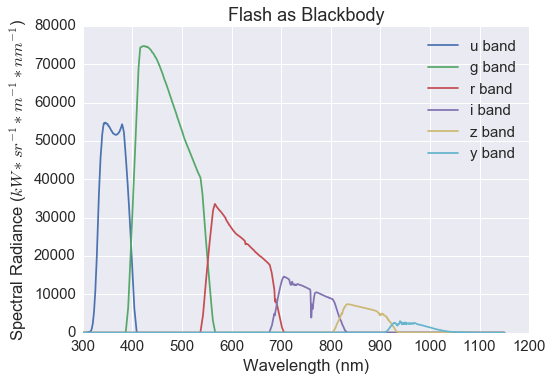}
   \includegraphics[width=0.45\textwidth]{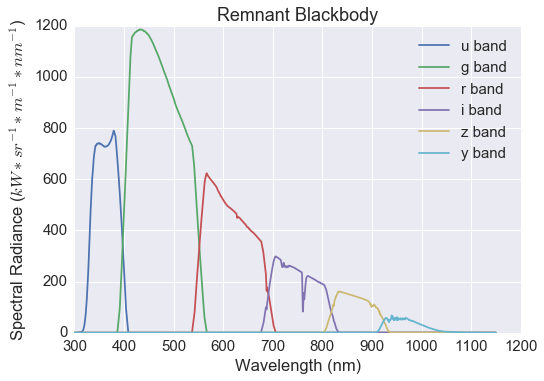}
   \includegraphics[width=0.45\textwidth]{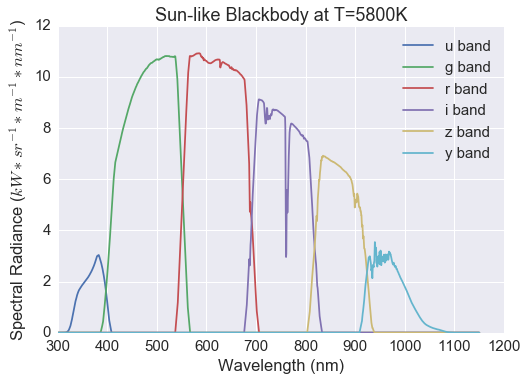}
  \end{center}
  \caption{The flash and remnant blackbody curves when convolved with the LSST filter throughputs. We include a Sun-like star as well for comparison.}
  \label{fig:convolve}
\end{figure*}

We treat the solar-mass star as our reference for calibrating the absolute magnitudes by using the method for determining the absolute magnitude in each band using the method that was outlined in \citet{Lund2014}. We then compare the total flux in each bandpass for the Sun and for the flash and remnant in order to get relative magnitudes, followed by absolute magnitudes. An important consideration here is that the radius of the planet must be included in these calculations, and so the remnant is a close analogue of a white dwarf in radius and temperature. The absolute magnitudes that we determine are listed in Table~\ref{table:abmag}. It becomes readily apparent that the remnant is generally no more than 1\% of the brightness of a solar-mass star, and the flash is only brighter than a solar-mass star in the \emph{u} band.

\begin{table}[H]
\caption{Absolute Magnitudes}
\label{table:abmag}
\centering
\begin{tabular}{llll}
\toprule
Band & Sun & Flash & Remnant \\
\midrule
\emph{u} & 7.23 & 6.53 & 11.16 \\
\emph{g} & 5.86 & 6.56 & 11.00 \\
\emph{r} & 4.49 & 6.23 & 10.52 \\
\emph{i} & 4.33 & 6.67 & 10.87 \\
\emph{z} & 4.29 & 6.98 & 11.13 \\
\emph{y} & 3.70 & 6.62 & 10.73 \\
\bottomrule
\end{tabular}
\end{table}

These results are even more constraining than they may appear at first glance. The simulated flash duration is 2 seconds, however LSST will have exposures that are 15 seconds in duration. To correctly get the measured apparent magnitude, this difference in duration has to be accounted for, and the flash will look on order of 2 magnitudes fainter in the 15-second exposures of LSST, meaning that it will be slightly fainter than the star. As an inhabited Earth-analog planet (and, therefore, any planet likely worth destroying) would be expected to be around a solar-mass star, the light from the flash and remnant would have to be of considerable brightness with respect to the host star to be observed, and it does not appear that this is the case.

There are, however, three scenarios that may result in the destruction event still being detectable. The first is if the star and planet are close enough to our Solar System that the planet's destruction can be angularly resolved. Given that LSST will saturate at 16th magnitude, however, it seems extremely unlikely that any geometry exists where this would be possible. The second is if the planet is orbiting a smaller star. A red dwarf, for example, will be several magnitudes fainter, particularly on the bluer end of the LSST filter set. In this case, the flash, and possibly the remnant, will be brighter than the host star. While red dwarfs have not been the typical stars searched for planets in the past, there is no reason to think that an inhabited planet could not orbit around a red dwarf. Finally, the flash in the \emph{u} band is still brighter than a solar mass star if it is observed instantaneously. In the case of LSST or other survey, this could also be accomplished by having a shorter exposure time, and so an exposure of 2-3 seconds would mean that any flash from a planetary explosion will be significantly brighter than the host star. In the case of LSST, however, the costs of this change to the observing schedule greatly outweigh this benefit as it would significantly curtail the observations that LSST will be able to make of fainter objects.

\section{Summary}

Astronomy has a relatively unique pattern of discovery when compared to other sciences in that much of astronomy is simply collecting large amounts of data with the hope that interesting and novel objects are discovered this way. Many major astronomical discoveries were not necessarily the results of pointed searches, but rather the luck of observing in the right place at the right time. This has certainly been the case for discoveries of astrophysical events, but may well be the case for observations of events linked to alien civilizations also. In this paper we have briefly explored the ability of the Large Synoptic Survey Telescope to observe an alien megastructure destroying a terrestrial planet. While it does not appear that LSST (or indeed, most telescopes) would be able to detect this for a terrestrial planet around a solar-mass star, there is a new hope that such an event would be easily observable in the case of the destruction of a terrestrial planet that orbits around a red dwarf.


\bibliographystyle{apalike}
\bibliography{ref_list}


\end{multicols}

\end{document}